\documentclass[aps,pra]{revtex4}
\usepackage{graphicx}

\begin{document}

\title{Three-party qutrit-state sharing}
\author{Zhang-yin Wang and Zhan-jun Zhang$^*$\\
{\footnotesize School of Physics \& Material Science, Anhui
University, Hefei 230039, China} \\
{\footnotesize $^*$Corresponding author. Email address:
zjzhang@ahu.edu.cn}}

\maketitle

\begin{minipage}{420pt}

{\bf Abstract} A three-party scheme for securely sharing an
arbitrary unknown single-qutrit state is presented. Using a general
Greenberger-Horne-Zeilinger (GHZ) state as the quantum channel among
the three parties, the quantum information (i.e., the qutrit state)
from the sender can be split in such a way that the information can
be recovered if and only if both receivers collaborate. Moreover,
the generation of the scheme
to multi-party case is also sketched.\\

\noindent {\it PACS numbers}: 03.67.Hk, 03.67.Dd, 03.65.Ud, 89.70.+c \\

\end{minipage}

\noindent {\bf I. Introduction}\\

Secret sharing was proposed firstly by Blakley et al[1] in 1979. It
can be depicted, in the simple case, as that a secret is divided by
a sender into two pieces for two receivers. The secret can be
reconstructed only if both receivers act in concert and neither of
them can get anything about the original message solely. In 1999,
this concept was generalized to quantum scenario by Hillery,
B\v{u}zek, and Berthiaume (HBB)[2]. They endowed it a novel concept
of quantum secret sharing (QSS). QSS is likely to play a key role in
protecting secret quantum information, e.g., in secure operations of
distributed quantum computation, sharing difficult-to-construct
ancilla states and joint sharing of quantum money, and so on. Hence,
after HBB's pioneering work, QSS, as an important branch of quantum
communication, has so far attracted a great deal of
attentions[3-28].

All the QSS works concentrate essentially on two kinds of problems.
One deals with the QSS of classical messages (i.e., bits) [2-14];
another deals with the QSS of quantum information [2, 15-28], where
the secret is an arbitrary unknown quantum state. The former is
usually referred as quantum secret sharing (QSS); the latter as
quantum state sharing (QSTS), which was first clearly termed by
Lance et al. in 2004[19]. As far as QSTS is concerned, the first
scheme was presented in 1999 by using a three-qubit or a four-qubit
Greenberger-Horne-Zeilinger (GHZ) state for securely sharing an
arbitrary unknown single-qubit state[2]. Soon later, Cleve et
al.[15] investigated a more general quantum ($k,n$) threshold QSTS
scheme. Bandyopadhyay[16] proposed a QSTS scheme using optimal
methods in 2000, and Hsu[17] proposed other QSTS scheme based on
Grover's algorithm in 2003. Recently, Li et al.[18] proposed an QSTS
scheme for sharing an unknown single-qubit state with a multipartite
joint measurement. Some QSTS schemes were implemented in cavity
QED[21-22]. Zhang et al.[23] proposed a multiparty QSTS of an
arbitrary unknown single-qubit state via photon pairs. Lance et
al.[24] proposed other continuous-variable QSTS scheme via quantum
disentanglement. Deng et al.[25-26] proposed two QSTS schemes for
sharing an arbitrary two-qubit state based on entanglement swapping.
Li et al.[27] proposed an efficient symmetric multiparty QSTS scheme
of an arbitrary $m$-qubit state with $m$ GHZ states. Very recently,
Gordon and Rigolin[28] proposed two new QSTS protocols where the
quantum channels are not maximally entangled states. Note that all
these QSTS protocols except for that in Refs.[18,24], however, only
treat single-particle {\it qubit} or multi-particle {\it qubit}
state. In this paper, we will propose a QSTS protocol for sharing an
arbitrary unknown single-particle {\it qutrit} state.

This paper is organized as follows. In section II, a three-party
QSTS scheme is presented by using quantum entanglement swapping,
and the scheme security is analyzed. In section III, the
three-party QSTS scheme is generalized to a multiparty case.
Finally, some summaries are given in section IV.\\

\noindent {\bf II. Three-party qutrit-state sharing scheme}\\

Suppose there are three legitimate users. Alice is the sender of
quantum information (i.e., an unknown qutrit state), Bob and Charlie
are two agents. Any agent can reconstruct Alice's quantum
information by collaborating with the other agent. Suppose Alice
owns qutrits 1 and 2, Bob qutrit 3 and Charlie qutrit 4. The
combined state of four particles is
\begin{eqnarray}
|\Phi\rangle_{1234}=|P\rangle_1\otimes|\psi\rangle_{234},
\end{eqnarray}
where
\begin{eqnarray}
|P\rangle_1=\alpha|0\rangle_1+\beta|1\rangle_1+\gamma|2\rangle_1,
\end{eqnarray}
\begin{eqnarray}
|\psi\rangle_{234}=\frac{1}{\sqrt{3}}(|000\rangle_{234}+|111\rangle_{234}+|222\rangle_{234}),
\end{eqnarray}
and $\alpha$, $\beta$ and $\gamma$ are complex and satisfy
$|\alpha|^2+|\beta|^2+|\gamma|^2=1$. Alice wants to send her
arbitrary unknown single-qutrit state $|P\rangle_1$ in such a way
that anyone of the two agents Bob and Charlie can reconstruct the
unknown state with the other's help. In order to achieve her goal,
Alice performs a generalized Bell-state projective measurement on
her qutrit pair (1, 2). After Alice's measurement, the system's
state evolves to one of the following nine possible results:
\begin{eqnarray}
|\Psi_{00}\rangle_{12}\langle\Psi_{00}|\Phi\rangle=\frac{1}{3}|\Psi_{00}\rangle_{12}(\alpha|00\rangle_{34}+\beta|11\rangle_{34}+\gamma|22\rangle_{34}),
\end{eqnarray}
\begin{eqnarray}
|\Psi_{01}\rangle_{12}\langle\Psi_{01}|\Phi\rangle=\frac{1}{3}|\Psi_{01}\rangle_{12}(\alpha|11\rangle_{34}+\beta|22\rangle_{34}+\gamma|00\rangle_{34}),
\end{eqnarray}
\begin{eqnarray}
|\Psi_{02}\rangle_{12}\langle\Psi_{02}|\Phi\rangle=\frac{1}{3}|\Psi_{02}\rangle_{12}(\alpha|22\rangle_{34}+\beta|00\rangle_{34}+\gamma|11\rangle_{34}),
\end{eqnarray}
\begin{eqnarray}
|\Psi_{10}\rangle_{12}\langle\Psi_{10}|\Phi\rangle=\frac{1}{3}|\Psi_{10}\rangle_{12}(\alpha|00\rangle_{34}+e^{-2\pi
i /3}\beta|11\rangle_{34}+e^{-4\pi i /3}\gamma|22\rangle_{34}),
\end{eqnarray}
\begin{eqnarray}
|\Psi_{20}\rangle_{12}\langle\Psi_{20}|\Phi\rangle=\frac{1}{3}|\Psi_{20}\rangle_{12}(\alpha|00\rangle_{34}+e^{-4\pi
i /3}\beta|11\rangle_{34}+e^{-8\pi i /3}\gamma|22\rangle_{34}),
\end{eqnarray}
\begin{eqnarray}
|\Psi_{11}\rangle_{12}\langle\Psi_{11}|\Phi\rangle=\frac{1}{3}|\Psi_{11}\rangle_{12}(\alpha|11\rangle_{34}+e^{-2\pi
i /3}\beta|22\rangle_{34}+e^{-4\pi i/3}\gamma|00\rangle_{34}),
\end{eqnarray}
\begin{eqnarray}
|\Psi_{21}\rangle_{12}\langle\Psi_{21}|\Phi\rangle=\frac{1}{3}|\Psi_{21}\rangle_{12}(\alpha|11\rangle_{34}+e^{-4\pi
i /3}\beta|22\rangle_{34}+e^{-8\pi i/3}\gamma|00\rangle_{34}),
\end{eqnarray}
\begin{eqnarray}
|\Psi_{12}\rangle_{12}\langle\Psi_{12}|\Phi\rangle=\frac{1}{3}|\Psi_{12}\rangle_{12}(\alpha|22\rangle_{34}+e^{-2\pi
i /3}\beta|00\rangle_{34}+e^{-4\pi i/3}\gamma|11\rangle_{34}),
\end{eqnarray}
\begin{eqnarray}
|\Psi_{22}\rangle_{12}\langle\Psi_{22}|\Phi\rangle=\frac{1}{3}|\Psi_{22}\rangle_{12}(\alpha|22\rangle_{34}+e^{-4\pi
i /3}\beta|00\rangle_{34}+e^{-8\pi i/3}\gamma|11\rangle_{34}),
\end{eqnarray}
where
\begin{eqnarray}
|\Psi_{nm}\rangle=\sum\limits_{j=0}^2e^{2\pi ij
n/3}|j\rangle\otimes|(j+m)mod3\rangle/\sqrt{3}, \ \  n \in
\{0,1,2\}, \ m \in \{0,1,2\}.
\end{eqnarray}
For each possible result, the treatment is similar in this paper. As
an enumeration, only one result is taken as an example hereafter.
Without loss of generality, suppose Alice's measurement result is
$|\Psi_{00}\rangle_{12}$. In this case, the qutrits 3 and 4 collapse
to the entangled  state
\begin{eqnarray}
|K_1\rangle_{34}=\frac{1}{3}(\alpha|00\rangle_{34}+\beta|11\rangle_{34}+\gamma|22\rangle_{34}).
\end{eqnarray}
This state can be rewritten as
\begin{eqnarray}
|K_1\rangle_{34}&=&\frac{1}{3}(\alpha|00\rangle_{34}+\beta|11\rangle_{34}+\gamma|22\rangle_{34})\nonumber\\
&=&\frac{1}{3}[\frac{1}{\sqrt{3}}|\xi_0\rangle_3(\alpha|0\rangle_4+\beta|1\rangle_4+\gamma|2\rangle_4)\nonumber\\
&+&\frac{1}{\sqrt{3}}|\xi_1\rangle_3(\alpha|0\rangle_4+e^{-2\pi i
/3}\beta|1\rangle_4+e^{-4\pi i /3}\gamma|2\rangle_4)\nonumber\\
&+&\frac{1}{\sqrt{3}}|\xi_2\rangle_3(\alpha|0\rangle_4+e^{-4\pi i
/3}\beta|1\rangle_4+e^{-2\pi i /3}\gamma|2\rangle_4)],
\end{eqnarray}
where
\begin{eqnarray}
&&|\xi_0\rangle=\frac{1}{\sqrt{3}}(|0\rangle+|1\rangle+|2\rangle),\nonumber\\
&&|\xi_1\rangle=\frac{1}{\sqrt{3}}(|0\rangle+e^{2\pi
i/3}|1\rangle+e^{4\pi i /3}|2\rangle),\nonumber\\
&&|\xi_2\rangle=\frac{1}{\sqrt{3}}(|0\rangle+e^{4\pi
i/3}|1\rangle+e^{2\pi i /3}|2\rangle).
\end{eqnarray}
The three states $\{|\xi_t\rangle, t=0,1,2\}$ are related to the
computation basis vectors $\{|0\rangle, |1\rangle, |2\rangle\}$, and
form a complete orthogonal basis set of a single-qutrit Hilbert
space. After her measurement, Alice publishes her result
$|\Psi_{00}\rangle_{12}$ and assigns either Bob or Charlie (she
makes the choice at random) to measure his qutrit in the complete
orthogonal basis set proposed above. Without loss of generality,
suppose Bob is assigned. If Bob's measurement result is
$|\xi_0\rangle_3$, the qutrit 4 (in Charlie's possession) is
projected onto
$\alpha|0\rangle_4+\beta|1\rangle_4+\gamma|2\rangle_4$. This state
is exactly the original state $|P\rangle$. Now Charlie cooperates
with Bob to get his result over a public channel. With Bob's help,
Charlie reconstructs the original state with no unitary operation.
While Bob's measurement result is $|\xi_1\rangle_3$, the qutrit 4 is
projected onto $\alpha|0\rangle_4+e^{-2\pi i
/3}\beta|1\rangle_4+e^{-4\pi i /3}\gamma|2\rangle_4$. Charlie
reconstructs the original state $|P\rangle$ by performing an unitary
operation $U_1=\sum\limits_{j=0}^2 e^{2\pi ij/3}|j\rangle\langle j|$
under Bob's help. Similarly, if Bob's measurement result is
$|\xi_2\rangle_3$, the qutrit 4 is projected onto
$\alpha|0\rangle_4+e^{-4\pi i /3}\beta|1\rangle_4+e^{-2\pi i
/3}\gamma|2\rangle_4$. In this case, conditioned on Bob's classical
bit for his result Charlie recovers the original state $|P\rangle$
by carrying out an unitary operation $U_2=\sum\limits_{j=0}^2
e^{4\pi ij/3}|j\rangle\langle j|$. So far, we have demonstrated the
three-party qutrit state sharing scheme of an arbitrary unknown
single-qutrit state. Note that in the above scheme Alice assigns the
agent Bob to measure the qutrit 3 and the agent Charlie to
reconstruct the original state $|P\rangle$. While the agent who is
assigned by Alice to do a single-qutrit measurement is Charlie and
Bob is assigned to recover the quantum information. As the symmetry,
the above procedure is also feasible. Here we do not state it
anymore.

Now let's analyze the scheme security. We consider two kinds of
eavesdroppers.

(a) Outside eavesdropper: Suppose there is an illegitimate user
named Eve. She wants to gain the quantum information which Alice
lets Bob and Charlie share. To achieve her goal, she entangles an
ancilla with the quantum channel during the particle distribution
process. For this case, the security check of the present
three-party qutrit state sharing scheme is very similar to that of
the protocol proposed by Hillery et al.[2]. That is, the security
depends completely on whether the three legitimate users have
securely shared the entangled GHZ state which are taken as the
quantum channel. Here we briefly review the check method. The
legitimate user Alice choose randomly a single-qutrit measurement
basis (MB) $\{|\xi_t\rangle, t=0,1,2\}$ or $\{|0\rangle,
|1\rangle, |2\rangle\}$ to measure her qutrit. After her
measurement, Alice tells the other two legitimate users Bob and
Charlie which MB she has chosen for her qutrit. Bob and Charlie
use the same MB as Alice to measure their respective qutrit. Their
measurement outcomes should be strongly correlated. If there
exists an eavesdropper Eve in the quantum line, her operation will
of course introduce some disturbance which will cause some qutrit
errors. Thus, when the legitimate users publicly compare their
results, they will find some incorrelation which means that the
quantum channel is disturbed. Alternatively, there may exist an
eavesdropper Eve. In this case, the quantum sharing process is
aborted. Incidentally, in our scheme the qutrit GHZ state is
assumed to be safely shared among legitimate users. This can be
achieved using the quantum purification and distillation or
quantum repeater techniques if the quantum channel noise or
decoherence is taken into account[29-34].

(b) Inside eavesdropper: suppose one of two legitimate agents
(say, Bob) is dishonest. He wants to solely and safely recover
Alice's quantum information without any assistance from Charlie.
To achieve his goal, Bob captures the qutrit Alice sends to
Charlie and then sends Charlie a fake qutrit he has prepared
before. In this case, only when Alice designates him to
reconstruct the state, he can successfully get the state
$|P\rangle$ and avoid the security detection. However, if Alice
designates not Bob but Charlie to reconstruct the state, then the
state reconstructed by Charlie will differ from the state Alice
has sent. In this case, if Alice and Charlie publicly compare the
state, the eavesdropping can be disclosed. Hence, the success
probability for the dishonest Bob is only 50$\%$ in each run.
During the whole sharing process, if the amount of the check state
is large enough, then the dishonest Bob will be revealed. \\

\noindent {\bf III. Multi-party qutrit-state sharing scheme}\\

Now let us generalize the three-party qutrit state sharing scheme
to multi-party case. Suppose there are $N+1$ legitimate users.
Alice is the quantum information sender. By the way, the quantum
information is still given by the equation 1. The other $N$ users
are Alice's agents, named as Bob (1st agent), Charlie (2nd
agent),..., Zach ($N$th agent), respectively. All the legitimate
users have successfully shared in advance a general $(N+1)-$qutrit
GHZ state
\begin{eqnarray}
|\psi'\rangle_{23...(N+2)}=\frac{1}{\sqrt{3}}(|00...0\rangle_{23...(N+2)}
+|11...1\rangle_{23...(N+2)}+|22...2\rangle_{23...(N+2)}).
\end{eqnarray}
Qutrit 2 belongs to Alice, and qutrits 3, 4,...$(N+2)$ to Bob,
Charlie,...Zach, respectively. Similarly, in order to split her
quantum information into $N$ parts for her $N$ agents, Alice
performs a generalized Bell-state projective measurement on her
qutrit pair (1,2) and publishes her result. After the generalized
Bell-state projective measurement, the system's state evolves to one
of the following nine possible results:
\begin{eqnarray}
|\Psi_{00}\rangle_{12}\langle\Psi_{00}|\Phi'\rangle=\frac{1}{3}|\Psi_{00}\rangle_{12}(\alpha|00...0\rangle_{34...(N+2)}+\beta|11...1\rangle_{34...(N+2)}+\gamma|22...2\rangle_{34...(N+2)}),
\end{eqnarray}
\begin{eqnarray}
|\Psi_{01}\rangle_{12}\langle\Psi_{01}|\Phi'\rangle=\frac{1}{3}|\Psi_{01}\rangle_{12}(\alpha|11...1\rangle_{34...(N+2)}+\beta|22...2\rangle_{34...(N+2)}+\gamma|00...0\rangle_{34...(N+2)}),
\end{eqnarray}
\begin{eqnarray}
|\Psi_{02}\rangle_{12}\langle\Psi_{02}|\Phi'\rangle=\frac{1}{3}|\Psi_{02}\rangle_{12}(\alpha|22...2\rangle_{34...(N+2)}+\beta|00...0\rangle_{34...(N+2)}+\gamma|11...1\rangle_{34...(N+2)}),
\end{eqnarray}
\begin{eqnarray}
|\Psi_{10}\rangle_{12}\langle\Psi_{10}|\Phi'\rangle=\frac{1}{3}|\Psi_{10}\rangle_{12}(\alpha|00...0\rangle_{34...(N+2)}+e^{-2\pi
i /3}\beta|11...1\rangle_{34...(N+2)}+e^{-4\pi i
/3}\gamma|22...2\rangle_{34...(N+2)}),
\end{eqnarray}
\begin{eqnarray}
|\Psi_{20}\rangle_{12}\langle\Psi_{20}|\Phi'\rangle=\frac{1}{3}|\Psi_{20}\rangle_{12}(\alpha|00...0\rangle_{34...(N+2)}+e^{-4\pi
i /3}\beta|11...1\rangle_{34...(N+2)}+e^{-8\pi i
/3}\gamma|22...2\rangle_{34...(N+2)}),
\end{eqnarray}
\begin{eqnarray}
|\Psi_{11}\rangle_{12}\langle\Psi_{11}|\Phi'\rangle=\frac{1}{3}|\Psi_{11}\rangle_{12}(\alpha|11...1\rangle_{34...(N+2)}+e^{-2\pi
i /3}\beta|22...2\rangle_{34...(N+2)}+e^{-4\pi
i/3}\gamma|00...0\rangle_{34...(N+2)}),
\end{eqnarray}
\begin{eqnarray}
|\Psi_{21}\rangle_{12}\langle\Psi_{21}|\Phi'\rangle=\frac{1}{3}|\Psi_{21}\rangle_{12}(\alpha|11...1\rangle_{34...(N+2)}+e^{-4\pi
i /3}\beta|22...2\rangle_{34...(N+2)}+e^{-8\pi
i/3}\gamma|00...0\rangle_{34...(N+2)}),
\end{eqnarray}
\begin{eqnarray}
|\Psi_{12}\rangle_{12}\langle\Psi_{12}|\Phi'\rangle=\frac{1}{3}|\Psi_{12}\rangle_{12}(\alpha|22...2\rangle_{34...(N+2)}+e^{-2\pi
i /3}\beta|00...0\rangle_{34...(N+2)}+e^{-4\pi
i/3}\gamma|11...1\rangle_{34...(N+2)}),
\end{eqnarray}
\begin{eqnarray}
|\Psi_{22}\rangle_{12}\langle\Psi_{22}|\Phi'\rangle=\frac{1}{3}|\Psi_{22}\rangle_{12}(\alpha|22...2\rangle_{34...(N+2)}+e^{-4\pi
i /3}\beta|00...0\rangle_{34...(N+2)}+e^{-8\pi
i/3}\gamma|11...1\rangle_{34...(N+2)}).
\end{eqnarray}
This means that Alice can get anyone of the nine possible results.
Similar to the three-party case, without loss of generality we
only take one result as an enumeration hereafter. Suppose Alice's
measurement result is $\Psi_{00}\rangle_{12}$. In this case, the
state of the qutrits $3, 4, ..., (N+2)$ is
\begin{eqnarray}
|K_2\rangle_{34...(N+2)}=\frac{1}{3}(\alpha|00...0\rangle_{34...(N+2)}
+\beta|11...1\rangle_{34...(N+2)}+\gamma|22...2\rangle_{34...(N+2)}).
\end{eqnarray}
This state can be reexpressed as
\begin{eqnarray}
&&|K_2\rangle_{34...(N+2)}\nonumber \\
&=&\frac{1}{3}(\alpha|00...0\rangle_{34...(N+2)}
+\beta|11...1\rangle_{34...(N+2)}+\gamma|22...2\rangle_{34...(N+2)})\nonumber \\
&=&(\frac{1}{\sqrt{3}})^{N+1}\sum\limits_{l_1=0}^2\sum\limits_{l_2=0}^2...\sum\limits_{l_{m-1}=0}^2
\sum\limits_{l_{m+1}=0}^2...\sum\limits_{l_N=0}^2 \\
&\times
&[|\xi_{l_1}\rangle_3|\xi_{l_2}\rangle_4...|\xi_{l_{m-1}}\rangle_{m+1}
(\alpha|0\rangle_{m+2}+\beta e^{-2\pi iL/3}|1\rangle_{m+2}+\gamma
e^{-4\pi i
L/3}|2\rangle_{m+2})|\xi_{l_{m+1}}\rangle_{m+3}...|\xi_{l_{N}}\rangle_{N+2}]\nonumber,
\end{eqnarray}
where
\begin{eqnarray}
L=\sum\limits_{i=1}^{m-1} l_{i}+\sum\limits_{j=m+1}^N l_{j}.
\end{eqnarray}
Alice can assign any agent to reconstruct the unknown state. In
other words, anyone of the $N$ agents has the chance to
reconstruct the unknown state. After Alice's assignment, all the
other agents should perform some operations and then help the
assigned agent to reconstruct the state. Without loss of
generality, we assume Alice assigns the $m$th agent to reconstruct
her original state. According to the equation 28, after the other
$N-1$ agents' measurement the qutrit in the $m$th agent's
possession is left into $\alpha|0\rangle_{m+2}+\beta e^{-2\pi
iL/3}|1\rangle_{m+2}+\gamma e^{-4\pi i L/3}|2\rangle_{m+2}$. If
all the other agents collaborate with the assigned agent, he/she
can reconstruct the original state $|P\rangle$ in his/her qutrit
by performing the unitary transformation $U_3=\sum\limits_{j=0}^2
e^{2\pi ijL/3}|j\rangle\langle j|$. 

The security of the multi-party qutrit state sharing scheme is same
as the security of the three-party qutrit state sharing scheme: any
eavesdropping leads to the discrepancy between the state that Alice
sends and the state that legitimate user reconstructs. Thus an
eavesdropping can be detected by publicly comparing a subset of the quantum states. \\

\noindent {\bf IV. Conclusion}

In summary, an interesting work for quantum state sharing of an
arbitrary unknown single-qutrit state has been done. In this paper,
we take a general GHZ state as the quantum channel. The state sender
Alice performs a generalized Bell-state projective measurement and
publishes her measurement result. As the symmetry, anyone (the
receiver) of the $N$ agents can regenerate the original state when
he/she collaborates with the others. The other agents are required
to perform one single-qutrit measurement on their respective qutrit.
Conditioned on Alice's two classical bits, the receiver can
reconstruct the original state via an unitary transformation only
when he/she gets the other agents' help. In addition, we use
multi-qutrit entangled stat, instead of EPR pairs, as the entangled
quantum system, and it is more useful in
maintaining security. \\

\noindent {\bf Acknowledgements} \\
This work is supported by the National Natural Science Foundation
of China under Grant No.10304022, the science-technology fund of
Anhui province for outstanding youth under Grant No.06042087, the
general fund of the educational committee of Anhui province under
Grant No.2006KJ260B, and the key fund of the ministry of education
of China under Grant No.206063. \\

\noindent {\bf References}

\noindent[1] G. R. Blakley, in Proceedings of the American
Federation of Information Processing 1979 National Computer
Conference (American Federation of Information Processing,
Arlington, VA, 1979), pp.313-317; A. Shamir, Commun. ACM {\bf22},
612 (1979).

\noindent[2] M. Hillery, V. B\v{u}zek, and A. Berthiaume, Phys.
Rev. A {\bf59}, 1829(1999).

\noindent[3] A. Karlsson, M. Koashi, and N. Imoto, Phys. Rev. A
{\bf59}, 162 (1999).

\noindent[4] D. Gottesman, Phys. Rev. A {\bf61}, 042311 (2000).

\noindent[5] W. Tittel, H. Zbinden, and N. Gisin, Phys. Rev. A
{\bf63}, 042301 (2001).

\noindent[6] V. Karimipour, A. Bahraminasab, and S. Bagherinezhad,
Phys. Rev. A {\bf65}, 042320 (2002).

\noindent[7] H. F. Chau, Phys. Rev. A {\bf66}, 060302 (2002).

\noindent[8] S. Bagherinezhad and V. Karimipour, Phys. Rev. A
{\bf67}, 044302 (2003).

\noindent[9] G. P. Guo and G. C. Guo, Phys. Lett. A {\bf310}, 247
(2003).

\noindent[10] L. Xiao, G. L. Long, F. G. Deng, and J. W. Pan,
Phys. Rev. A {\bf69}, 052307 (2004).

\noindent[11] S. K. Singh, Phys. Rev. A {\bf71}, 012328 (2005).

\noindent[12] Z. J. Zhang, Phys. Lett. A {\bf342}, 60 (2005).

\noindent[13] Z. J. Zhang and Z. X. Man, Phys. Rev. A {\bf72},
022303 (2005).

\noindent[14] Z. J. Zhang {\it et al}., Phys. Rev. A {\bf71}, 044301
(2005); Chin. Phys. Lett. {\bf22} 1588 (2005).

\noindent[15] R. Cleve, D. Gottesman, and H. K. Lo, Phys. Rev. Lett.
{\bf83}, 648 (1999).

\noindent[16] S. Bandyopadhyay, Phys. Rev. A {\bf62}, 012308 (2000).

\noindent[17] L. Y. Hsu, Phys. Rev. A {\bf68}, 022306 (2003); L. Y.
Hsu and C. M. Li, Phys. Rev. A {\bf71}, 022321 (2005).

\noindent[18] Y. M. Li, K. S. Zhang, and K. C. Peng, Phys. Lett. A
{\bf324}, 420 (2004).

\noindent[19] A. M. Lance, T. Symul, W. P. Bowen, B. C. Sanders, and
P. K. Lam, Phys. Rev. Lett. {\bf92}, 177903 (2004).

\noindent[20] F. G. Deng {\it et al}., Phys. Lett. A {\bf337}, 329
(2005); F. G. Deng, X. H. Li, H. Y. Zhou, and Z. j. Zhang, Phys.
Rev. A {\bf72}, 044302 (2005).

\noindent[21] Y. Q. Zhang, X. R. Jin, and S. Zhang, Phys. Lett. A
{\bf341}, 380 (2005).

\noindent[22] Z. J. Zhang, Opt. Commun. {\bf261} 199 (2006).

\noindent[23] Z. J. Zhang {\it et al}., Eur. Phys. J. D {\bf33},
133 (2005).

\noindent[24] A. M. Lance et al., Phys. Rev. A {\bf71}, 033814
(2005).

\noindent[25] F. G. Deng {\it et al}., Phys. Rev. A {\bf72},
044301 (2005).

\noindent[26] F. G. Deng, X. H. Li, C. Y. Li, P. Zhou, and H. Y.
Zhou, e-print quant-ph/0509029.

\noindent[27] X. H. Li {\it et al}., J. Phys. B {\bf39}, 1975-1983
(2006).

\noindent[28] G. Gordon and G. Rigolin, Phys. Rev. A {\bf73},
062316 (2006).

\noindent[29] C. H. Bennett, D. P. DiVincenzo, J. A. Smolin, and
W. K. Wootters, Phys. Rev. A {\bf54}, 3824 (1996).

\noindent[30] C. H. Bennett, G. Brassard, C. Crepeau,  R. Jozsa,
A. Peres, and W. K. Wootters, Phys. Rev. Lett. {\bf70}, 1895
(1993).

\noindent[31] H. J. Briegel, W. Dur, J. I. Cirac, and P. Zoller,
Phys. Rev. Lett. {\bf81}, 5932 (1998).

\noindent[32] W. Dur, H. J. Briegel, J. I. Cirac, and P. Zoller,
Phys. Rev. A {\bf59}, 169 (1998).

\noindent[33] H. K. Lo and H. F. Chau, Science {\bf283}, 2050
(1999).

\noindent[34] S. J. van Enk, J. I. Cirac, and P. Zoller, Phys.
Rev. Lett. {\bf78}, 4293 (1997).

\enddocument